# How question quality drives Web performance in Community Question Answering sites

**Alyssa Shuang Sha**

Australian National University

LF Crisp Building 26C ACT, 2601

shuang.sha@anu.edu.au

**Yingnan Shi**

Australian National University & CSIRO

LF Crisp Building 26C ACT, 2601

yingnan.shi@anu.edu.au

**Armin Haller**

Australian National University

PAP Moran Building 26B ACT, 2601

armin.haller@anu.edu.au

## Abstract

*Users are posting millions of questions on Community Question Answering (CQA) sites each day. The quality of those questions significantly affects the satisfactions of the sites' users and, therefore, sites' traffic. We gathered 15 question-quality related features from one of the largest CQA sites and the site's page view data to estimate the scale of the effect in the corresponding time series. By using a Grey Relational Analysis, we rank those question quality features and estimate the relative strength of these factors on a page's view numbers. Our results show that question quality features have significant influence on Web performance. We generate a ranked list of features and find that questions' digital popularity and questioner-related features can drive the page traffic more than textual features and question difficulty. The implications of the findings for Web growth and future research are discussed.*

## Introduction

Community Question- Answering (CQA) systems, such as Yahoo! Answers, Stack Overflow, Quora and Zhihu, belong to a prominent group of successful and popular Web 2.0 applications which are used every day by millions of users to find answers on complex, subjective, or context-dependent questions. They have become some of the most valuable platforms to create, share, and seek a massive volume of human knowledge (Kuang et al., 2019). Questions play a critical role in CQA systems: A good question can attract users' initial attentions and motivates knowledge sharing intentions, which in turn can lead to an increase in answering attempts, a quicker response rate and higher answers' quality (Srba & Bielikova, 2016). However, it has been shown that the degree of question quality varies greatly. For example, in Yahoo! Answers, a few questions acquired thousands of tag-of-interests and answering attempts, while most questions accumulated none at all (Li et al., 2012).

The quality of questions in a CQA site not only affects its users' subsequent answering intentions and behaviours, but also influences the satisfaction level of the knowledge seekers who account for a substantial level of incoming traffic (Q. Liu et al., 2011; Y. Liu et al., 2008). This paper is in line with studies on the question quality and CQA performance. In particular, we aim to examine how quality features of a question can drive its Web performance. We started with a systematic review and integration of the existing question quality measurement models (see Figure 1). We then, based on the model we have created, crawled data, including questions' quality features, from Zhihu, a Quora-like platform, which is the largest CQA social platform. We use Deng's Grey Relational Analysis model (GRA) to rank questions' quality features and estimate the relative strength of these factors on a page's view numbers. GRA is suitable for solving the complicated interrelationships between multiple factors and variables. Compared to other regression methods, GRA, which has been largely applied to performance evaluation and factor effect evaluation, can handle both incomplete information and unclear problems very precisely (Morán et al., 2006). By adopting this method, we estimated and ranked relationships between the question features and the corresponding pageviews. Pageviews, whilst being only one of many measurable attributes of Web traffic to a website, has been shown to be a good measure for Web performance (Z. Liu et al., 2001). Based on the correlation degree of all factors to pageview numbers which we used as a proxy to measure Web performance, we rank those question quality factors and explore which factors can exert a relatively more substantial impact on Web performance.

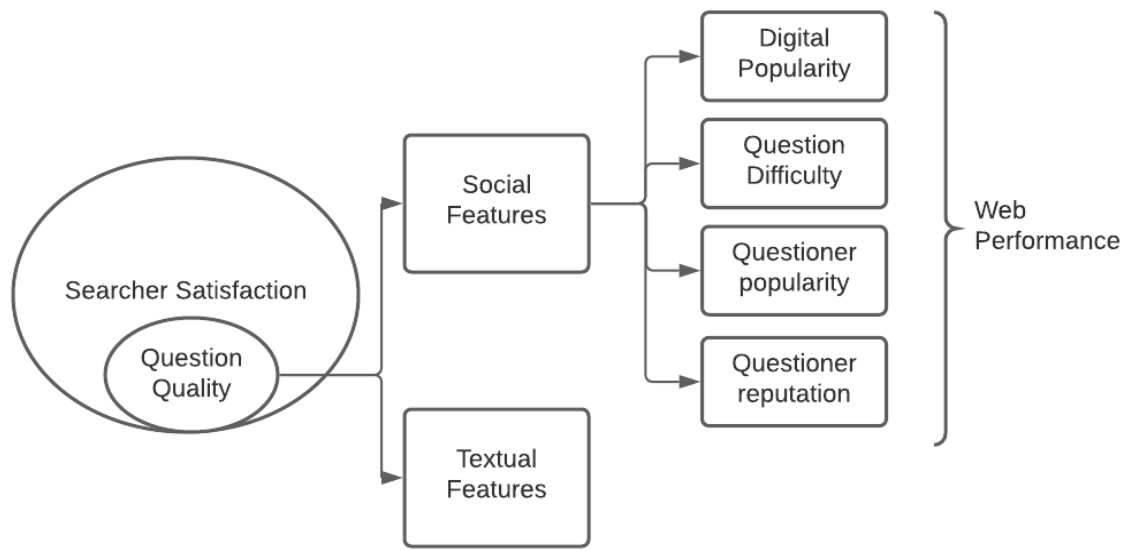

**Figure 1. Structure of question quality study in CQA**

Previous researchers have studied how to rank answers; however, few studies have been performed to rank features that measure the quality of questions. How strong the different features of the question quality affect website performance still remain unclear. Thus, this paper tries to comprehensively measure the question quality in CQA and build the connection between a question's quality features and Web performance. This research adopted the GRA model and ranks those question quality features based on the strength of various features on Web performance.

The remainder of this paper is structured as follows. In Section 2, we review the related work and analyze potential research gaps. We then present the experimental data and the regression model in Section 3. In Section 4, we present the results of our study and meaningful findings. In Section 5, we conclude the paper and discuss some limitations and potential future research directions.

# Literature review

### Approaches concerned with question quality assessment
Question quality assessment refers to the estimation of questions' quality in order for them to be distinguished/classified as high- and low-quality questions. "High quality" questions are supposed to attract increased user attention, more answer attempts and receive better answers within a short period. Otherwise, questions failing to achieve the three criteria are labeled as "low quality" questions since these questions neither meet user needs nor contribute to the knowledge base of the community (Li et al., 2012).

Many factors decide the quality of documents (or contents). Strong et al. (1997) listed 15 factors and classified those factors into 4 categories: contextual, intrinsic, representational and accessibility. Zhu and Gauch (2000) came up with 6 factors including currency, availability, information, authority, popularity and cohesiveness to define the quality metrics of Web pages. A comprehensive research has been undertaken by Srba and Bielikova (2016), where they covered various research issues of CQA site since 2016, including a question's quality measurement. Several works investigated questions' textual quality. Agichtein et al. (2008) used textual features such as length, number of unique words, word overlap between question and answer to identify the best answer. Suzuki et al. (2003) investigated how various contextual information included in a question can lead to better answers. Yeniterzi and Callan (2014) proposed a more topic-focused authority network construction approach which adopted textual features in social media. This research provides topic-specific authority modelling of users. Bhat et al. (2015) combined label-related features and statistical (textual) features of the question content (such as the title length of the question, the length of the post, the number of code fragments, etc.), and then used machine learning methods such as logistic regression, support vector machines and decision trees to predict whether the question can be answered before a certain time.

CQA services are rich in metadata such as user ratings, user comments, and users' points (Cai et al., 2018). In recent studies, researchers tap into these social features together with textual features extracted from surface textual content to identify quality answers from a CQA corpus.

Jeon (2006) integrated social features such as askers' authority and users' endorsement into a ranking framework to retrieve answers that are relevant, factual and of high quality. Ponzanelli (2014) proposed a system to identify low-quality questions on Stack Overflow. They made a classification system using both textual and social features. Li et al. (2012) applied the term "question quality" to represent the question's "social quality", which involves three dimensions (1) user attention; (2) answering attempts; and (3) best answer. In Yao's study, textual features and askers' reputation have been selected for question quality predictions (Yao et al., 2013). Pradeep K. Roy et al. (2018) characterized reputation collectors based on their answering activity, type of question answered and reputation points against those questions. He adopted several features such as vote-up amount, and accepted answers amount to capture questioners' reputation and divided them into different dimensions. Question difficulty was taken into consideration by Hanrahan (2012) who developed indicators for hard problems and experts. Huna et al. (2016) use the needed time to answer a question to measure question difficulty as a part of a novel reputation mechanism that focuses primarily on the quality and difficulty of users' contributions. Some studies further construct a Multi-feature-based Question-Answerer Model (MQAM) which adopts characteristics of answerers (e.g., the answerer's interest, professional level, active level) and questions (e.g., question category and difficulty) to predict users' activity in CQA (Burlutskiy et al., 2016; YueLiu et al., 2019). A comprehensive scale that Liu et al. (2011) proposed to measure searcher's satisfaction has been adopted by many researchers, which added a question-answer relationship as a new factor besides question's textual features, asker-user history and category features. However, Ho et al. (2020) aimed to avoid social influence biases arising from these indicators by predicting the quality from a semantic evaluation of the question text.

Previous studies of community question answering sites have mainly considered first-order effects, namely, the satisfaction of the original question asker with the posted answers. However, CQA sites have significant secondary benefits. Previously answered questions are likely to be useful for future askers of similar or related questions (Koufaris, 2002) and may alleviate the need of a user asking a question in the first place. Indeed, global Web search engines create substantial incoming traffic for Q&A sites through bringing secondary benefits, which can provide sustainable advertising revenue for the platform. Y. Liu et al. (2008) identified three key characteristics of a searcher's satisfaction, namely, query clarity, query-question match, and answer quality. It can be seen that questions' quality, as part of a searcher's satisfaction, has a significant impact on Web performance.

### Web Performance
The use of Web performance here refers to the subjective evaluation and judgment of a specific website by network users. Web performance is the key outcome variable in Web experiences that a website can provide to consumers. Behavioral outcomes such as technology acceptance or computer usage, are results of performance evaluations (Huang, 2005). Web traffic as an important indicator for a website's performance evaluation can provide useful information for users to estimate a site's validation and popularity (Z. Liu et al., 2001; Stephen, 2019). Quantitative methods focus on the performance measurement of a website whereas qualitative methods estimate the user's opinion of a website (Prantl & Prantl, 2018). The three common theoretical frameworks that have been employed for Web site evaluation include: technology acceptance model (TAM), flow theory, and human computer interaction (HCI) (Koufaris, 2002). Many different methods have been proposed to automate Web analysis so as to provide insights into the status of a website's usage. Ansari and Gadge (2012) proposed 18 factors to check a website's trustworthiness, including its Alexa rank and Alexa inbound links. Quoniam's (2011) approach for estimating website traffic is based on the position of a measured website in a search engine's results page.

In this paper, we adopted the visit number of each question, or in other words, the view number of the page where the question is located, as a manifestation of performance for the Web page. In order to reduce the impact of traffic fluctuations on the entire Q&A platform on the page views of each question, we use the monthly traffic of the platform as a control variable.

SEMRush is a well-known platform that applies this approach to determine visitor traffic from search engines (Prantl & Prantl, 2018). SEMrush uses Google Analytics data to examine paid advertisement strategies, keyword grouping and management, price-per-click effectiveness, and search engine optimization of websites hosted around the world. Many recent studies have used website traffic and ranking data on SEMRush to conduct research on different topics. Taylor and Bicak (2020) used SEMRush data to detect the amount of web traffic generated by paid adwords per month. Čižinská et al. (2016) extracted real Web traffic data provided by SEMRush to predict the relative value of the digital firm. The traffic analysis tool of advanced SEMRush provide absolute values for each website and also represent the traffic grouped by geographical distribution and device types. Besides, compared with other websites, SEMRush has more historical traffic data which can be traced back to January 2017. Due to the above advantages, we chose the database from SEMRush as the source of our monthly traffic data.

**Grey Relational Analysis model**
In a complex and multivariate time series system, many factors simultaneously influence the system to determine its state of development. When the relationship between various factors is 'grey' which means unclear, uncomplete and uncertain (Kung & Wen, 2007; Morán et al., 2006; Wen, 2004), it is hard to decide which factor has a higher influence on the system. The Grey Relational Analysis (GRA) as a statistical model was developed by Professor Deng Julong from Huazhong University of Science and Technology, People's Republic of China (Julong Deynrt, 1989). GRA is employed to search for a grey relational grade (GRG), which is an effective tool that can be used to make a system analysis, describe the relationships among the factors and to determine the important factors that significantly influence some defined objectives. Compared to regression analysis and factor analysis, a grey relational analysis can adopt small samples that do not necessarily follow a normal distribution to measure the correlation degree of factors (S. Liu et al., 2013). GRA, thereby, proposes a ranking scheme that ranks the order of the grey relationship among dependent and independent factors, which allows us to decide which features need to be considered to drive Web performance more precisely.

In summary, we found question quality is an important part in CQA research, yet we know little about which quality features play a more critical role in boosting a website's performance. Specifically, we found no research that analyses how the question quality influences Web performance based on a time series analysis. This research fills this gap in CQA research on how the question quality influences website traffic. It can guide researchers and platform owners on how to improve Web performance and user engagement in CQA.

# Approach

## Data description and measurement definition

**Zhihu dataset**
Using an API (api.zhihu.com) and Python zhihu_oauth package (Pypi, 2019), we collected data on the questions' digital popularity, questioner-related features, question difficulty and questions' textual features from January, 2017 to September, 2020. Questions on this site can be retrieved using a Question ID (QID) as a key, so we generated random numbers to randomly retrieve a number of questions. Based on our observation, we found that the question IDs are always eight to nine digits long and the absolute values of the QIDs monotonically increase with time. We hence traced the QID for the first question that was asked on October 1st, 2017 and ascertained that it should have a QID that is not smaller than 66,080,000; The last question that was asked on September 30th, 2020 should be no larger than 423,670,000. Hence, we generated random numbers within a range of [66080000, 423670000].

Invalid random QIDs were automatically filtered out as they would cause a 404 error. Once a valid QID has appeared, our program recorded the question's (if any) contextual and browsing information (including submission time, content, and followers count), its corresponding answers' information (submission time, content, and upvotes received), and questioners' contextual and social information (including accumulated follower and upvote count, badges). 63,974 valid questions were retrieved.

Also using Python, we then developed a selenium-based crawler (Pypi, 2018) to capture pageview data for each valid QID. We performed several preprocessing steps to remove "noisy" records. We removed all outliers for features with an absolute value greater than one that were beyond its Mean +/- 3*SD. A 'None' value was assigned to fields corresponding to answers' information if the question receives no answers. Lastly, all data concerning the independent variables were normalized before the regression analysis was conducted.

**General website traffic**
SEMRush offers a Traffic Analysis metric, such as numbers of visits and user engagement metrics, for a wide range of websites, including zhihu.com. We retrieved the total Web traffic data for the Zhihu platform including desktop and mobile data from SEMRush since January 2017. Finally, we describe our data in Table1.

|  | Mean | SD | Maximum | Minimum |
|---|---:|---:|---:|---:|
| Pageview | 4370.92 | 60879.78 | 4995891 | 1 |
| Follower count | 6.96 | 41.41 | 2103 | 0 |
| Comment count | 0.10 | 0.69 | 20 | 0 |
| Answer count | 3.05 | 10.70 | 299 | 0 |
| Questioner follower count | 230.06 | 17102.48 | 3316528 | 0 |
| Questioner answers count | 22.68 | 91.12 | 2177 | 0 |
| User endorsement (vote up) | 341.20 | 3971.63 | 229601 | 0 |
| User endorsement (thanks) | 49.25 | 551.47 | 25815 | 0 |
| Badges | 0.01 | 0.07 | 1 | 0 |
| Question title length | 21.47 | 11.04 | 51 | 3 |
| URL count | 0.43 | 1.61 | 48 | 0 |
| Question detail length | 124.36 | 316.29 | 8862 | 0 |
| Question and best answer length ratio | 1.64 | 10.25 | 874 | 0 |
| Wh type word ratio | 0.00 | 0.00 | 0 | 0 |
| Number of non-stop word overlap/length | 0.03 | 0.05 | 1 | 0 |
| Textual Features | 147.93 | 321.53 | 8916.01 | 3 |
| Digital popularity | 10.11 | 49.36 | 2412.00 | 0 |
| Questioner Reputation | 413.14 | 4509.13 | 256717.00 | 0 |
| Best answer received time (Days) | 52.75 | 162.68 | 1346.91 | 0 |

**Table 1. Descriptive Statistics for the Zhihu Dataset**

In order to carry out the relational regression analysis with the pageviews of the corresponding month, we calculated the mean value of the relevant parameters of 2,000 questions each month as our independent variables which will be the input to the GRA model.

**Question quality measurement scale**

Our measurement features are organized around the entities in a question answering community about question quality: social features and textual features. We now review the features we have used to represent our problem. The complete list is reported in Table 2.

| Social Features (8 total) | | Description |
|---|---|---|
| **#Digital popularity** | | |
| S_A: Follower count | | Total number of followers for this question |
| S_A: Comment count | | Number of comments added by other users |
| S_A: Answer count | | Number of total answers received for this question |
| **#Question difficulty** | | |
| S_S: Best answer received time | | Question posting time minus best answer posting time |
| **#Questioner's popularity** | | |
| S_QR: Questioner follower count | | Number of followers of the questioner |
| **#Questioner's reputation** | | |
| S_QR: User endorsement | Thanks | Number of 'thanks' received from other users |
| | Vote up | Number of vote up from other users |
| S_QR: Badges (yes=1 no=0) | | If the questioner has been awarded a badge, the value is one; otherwise, the value is zero |
| S_QR: Questioner answer count | | Number of all answers posted by this questioner on the platform |

| Textual Features (6 total) | Description |
|---|---|
| T: Question content length | Number of characters in question content |
| T: Question title length | Number of words in question title |
| T: Question and best answer length ratio | The ratio of the length of the question divided by the length of the best answer |
| T: Number of non-stop word overlap/length | The ratio of nonstop words and the length of the question content |
| T: Wh-type word ratio | The ratio of wh-word introducing the question title divided by the length of the question |
| T: URL count | Number of links in the question |

**Table 2. Measurement scale of question quality on Zhihu**

a. **Social Features:**

We divided this group into digital popularity/question popularity, question difficulty, questioner's popularity and questioner's reputation. Digital popularity of branded content is most commonly conceptualized as the number of likes, brand-related comments, and shares performed by online users which reflect users' attention (Karpinska-Krakowiak & Modlinski, 2020). We applied question's followers count, number of answer attempts and number of comments to measure the question's ability to attract user attention. In order to capture the question difficulty, we adopted Hanrahan's method (2012) and chose to use the best answer received period to detect whether the problem has been resolved in a timely and effective manner.

Since question quality is, to a large extent, reliant on asker's activity history (Y. Liu et al., 2008), we find that a questioner's reputation is unique to question-answering communities, and particularly crucial for our study. Thus, we decided to use a questioner's follower count as their popularity, user's endorsement, questioner's badges and questioner's total answer number as questioners' reputation to measure question quality in CQA.

**b. Textual Features:**

This group includes traditional question answering features such as the wh-type (e.g., "what" or "where") words ratio, the length of the subject (title) and detail (description) of the question. We derive word n-gram (unigram and bigram) features from the text of the question as more specific features to communities, which include the URL count in question content, question and best answer length ratio, and non-stop word ratio.

## Statistical method

The first step to do a grey relational analysis is the data pre-processing to transfer the original data sequence into a comparative sequence. There are two processes involved in this step: data representation and data normalization. In this study, our original data series ($X_0$) represents the pageviews. The mean value of features each month are our comparative series ($X_i$). The expectancy of our study is the higher-the-better, which means that if the value of the grey relational grade is higher, than there is a strong relationship between the comparative and the reference series. The data normalization can be expressed by:

$$x^*(k) = \frac{x_i^0(k) - \min x_i^0(k)}{\max x_i^0(k) - \min x_i^0(k)}$$

Where i = 1.....m ; k = 1.....n. m is number of experimental data items, n is the number of parameters; $x_i^0(k)$ is the original sequence, $x_i^*(k)$ is the sequences after data pre-processing, $\min x_i^0(k)$ and $\max x_i^0(k)$ are the smallest and the largest value of $x_i^0(k)$.

The second step is to locate the grey relational coefficient by using the formula below:

$$\delta_i(k) = \frac{\Delta min + \delta \Delta max}{\Delta_{0,i}(k) + \delta \Delta max}$$

where, $\Delta_{0,i}$ = deviation sequences of the reference sequence and comparability sequence $x_0^*(k)$ = the reference sequence, and $x_i^*(k)$ = the comparative sequence. $\delta$ is known as identification coefficient with $\delta \in [0,1]$. In our study, $\delta = 0.5$ is adopted because it offers a moderate distinguishing effect and stability (Fu et al., 2001).

After the grey relational coefficient is derived, the grey relational grade (GRG) is calculated by averaging the value of the grey relational coefficients. The grey relational grade can be calculated using the formula below:

$$\gamma_i = \frac{1}{n}\sum_{k=1}^{n}\delta_i(k)$$

where $\gamma$ represents the correlation degree between the reference sequence and comparative sequence. In this study, a GRG (correlation degree) is used to indicate the relationship between Zhihu's Web traffic indicator (reference sequence) and the question quality features (comparability sequence). Therefore, if a particular comparability sequence is related closer than the other comparability sequence to the reference sequence, the GRG of that comparability sequences with the reference sequence will be higher. For instance, if $\gamma(x_0, x_1) > \gamma(x_0, x_2)$, the feature $x_1$ is closer to the dependent variable ($x_0$) than the feature $x_2$. Generally, $\gamma_i > 0.9$ indicates a marked influence, $\gamma_i > 0.8$ indicates a relatively marked influence, $\gamma_i > 0.7$ indicates a noticeable influence and $\gamma_i < 0.6$ indicates a negligible influence (Fu et al., 2001).

*How question quality drives Web performance in CQA*# Results

Table 3 indicates that the grey relational grade between all the features of a question's quality and page traffic. In the table, the different factors that affect the question quality are ranked according to the degree of relevance to the page traffic. The top-ranked features are the digital popularity-related features, followed by a series of questioner-related features which capture a questioner's popularity and a questioner's reputation. However, the textual features are at the bottom of the ranking list.

The results are grouped into different categories in Table 4. According to this table, digital popularity and a questioners' reputation are ranked slightly higher than textual features and question difficulty. In general, the results indicate that all the categories of the features have a markable influence on the pageviews. Platform traffic as a control factor has also been added into our regression model. The results show that the control variable does not have a significant impact on the statistical results.

| Rank | Features ($X_i$) | Grey Relational Grade ($\gamma_i$) |
| --- | --- | --- |
| 1 | Follower count | 0.970055 |
| 2 | Comment count | 0.96664 |
| 3 | Answer count | 0.957914 |
| 4 | User endorsement (vote up) | 0.957771 |
| 5 | Questioner answer count | 0.955894 |
| 6 | User endorsement (thanks) | 0.954294 |
| 7 | Questioner follower count | 0.936614 |
| 8 | Title length | 0.934529 |
| 9 | Badge | 0.933669 |
| 10 | URL count | 0.931183 |
| 11 | Question detail length | 0.930063 |
| 12 | Best answer received time (Days) | 0.929469 |
| 13 | Question and best answer length ratio | 0.921683 |
| 14 | Nonstop word ratio | 0.920797 |
| 15 | Wh-type work ratio | 0.902555 |

**Table 3. Grey relational analysis result of question quality features**

| Feature Group | Grey Relational Degree |
| --- | --- |
| Digital popularity | 0.946191 |
| Questioner reputation | 0.926154 |
| Questioner popularity | 0.916521 |
| Textual Features | 0.910642 |
| Question difficulty | 0.857626 |

**Table 4. Grey relational analysis result of question quality feature groups**

# Conclusion and Discussion

In this paper, we have conducted an empirical study based on a dataset that we retrieved from crawling Zhihu to investigate the relationship between question quality and the page traffic on the Q&A site. Based on previous research on question quality measures in CQA, we have produced a comprehensive measurement scale that includes social and textual features. By producing a ranked list, this study shows that all the features of question quality have a crucial positive impact on Web performance. Our results revealed that besides the question popularity features, questioner reputation features have more

AIS Special Interest Group on IS/IT in Asia Pacific Workshop, Hyderabad 2020    8

significant influence on Q&A page traffic than the textual features like length of title and content, non-stop word ratio, wh-word ratio or number of links. Compared to digital popularity, a questioners' reputation and textual features, question difficulty has relatively less impact on pageviews.

We experimented with a robust real-world dataset and adopted the GRA model to generate a ranking list for all the features and their groups, filling the gaps in previous studies on the complicated relationship between question quality and pageviews. We have shown the importance of question quality to Web performance in CQA. Our study opens a promising direction towards modelling the Web traffic growth process via the GRA ranking system in CQA, resulting in practical improvements to a Q&A platform's Web performance. As we have mentioned in the literature review, there are many quality assessments features and Web performance measurements, but we only considered a fraction of them. In future work, we will build a more comprehensive model to include more quality features and investigate their relationships and relative impacts on Web performance.

There are some limitations to our study. Firstly, we did not crawl several essential features such as 'like' and 'favorited' numbers for each question on Zhihu. To collect information on a questioner, we filter out those anonymously asked questions that can also have marked influence and possibly result in slight errors in the statistical results. Our study is also based on the assumption that the question-answer matching mechanism has not changed significantly since 2017 in Zhihu. Based on our current knowledge, there might be several external shock events, such as the introduction of for-paying questions, exclusive questions, paid live sessions etc., that happened in Zhihu which may have some impact on users' question-answering behaviours and relevant matching mechanisms. Therefore, a future research direction is to collect panel data and capture the potential relationship between the platform's new features from external shock events and Web performance.

Also, current results lead us to further explore other global CQA platforms and conduct a comparative analysis and see whether the same relationships exist in similar communities. We also plan to add classifiers into this research to study the detailed relationship among different topics. Because our research structure's base is searcher's satisfaction which including other feature groups like answers' quality and question-answer matching systems, we consider adopting the same ranking method and setting the ground truth on other feature groups to pave the way for the development of a more robust CQA service.

# References

Agichtein, E., Castillo, C., Donato, D., Gionis, A., & Mishne, G. (2008). Finding High-quality Content in Social MediaAgichtein, Eugene Castillo, Carlos Donato, Debora Gionis, Aristides Mishne, Gilard. *International Conference on Web Search and Data Mining*, 183–193. https://doi.org/10.1145/1341531.1341557

Ansari, S., & Gadge, J. (2012). Architecture for Checking Trustworthiness of Websites. *International Journal of Computer Applications*, *44*(14), 22–26. https://doi.org/10.5120/6332-8706

Bhat, V., Gokhale, A., Jadhav, R., Pudipeddi, J., & Akoglu, L. (2015). Effects of tag usage on question response time: Analysis and prediction in StackOverflow. *Social Network Analysis and Mining*, *5*(1), 1–13. https://doi.org/10.1007/s13278-015-0263-3

Burlutskiy, N., Fish, A., Ali, N., & Petridis, M. (2016). Prediction of users' response time in Q&A communities. *Proceedings - 2015 IEEE 14th International Conference on Machine Learning and Applications, ICMLA 2015*, 618–623. https://doi.org/10.1109/ICMLA.2015.190

Cai, S., Luo, Q., Fu, X., & Ding, G. (2018). Paying for Live Broadcast : Predicting Internet Knowledge Product Sharing. *CONF-IRM 2018 Proceedings*, 25.

Čižinská, R., Krabec, T., & Venegas, P. (2016). FieldsRank: The Network Value of the Firm. *International Advances in Economic Research*, *22*(4), 461–463. https://doi.org/10.1007/s11294-016-9604-x

Fu, C., Zheng, J., Zhao, J., & Xu, W. (2001). Application of grey relational analysis for corrosion failure of oil tubes. *Corrosion Science*, *43*(5), 881–889. https://doi.org/10.1016/S0010-938X(00)00089-5

Hanrahan, B. v., Convertino, G., & Nelson, L. (2012). Modeling problem difficulty and expertise in StackOverflow. *Proceedings of the ACM Conference on Computer Supported Cooperative Work, CSCW*, 91–94. https://doi.org/10.1145/2141512.2141550

Ho, M. K., Tatinati, S., & Khong, A. W. H. (2020). Distilling Essence of a Question : A Hierarchical Architecture for Question Quality in CQA Sites. *2020 International Joint Conference on Neural Networks*, *1*, 1–7.

Huang, M. H. (2005). Web performance scale. *Information and Management*, *42*(6), 841–852. https://doi.org/10.1016/j.im.2004.06.003

Huna, A., Srba, I., & Bielikova, M. (2016). Exploiting content quality and question difficulty in CQA reputation systems. *Lecture Notes in Computer Science (Including Subseries Lecture Notes in Artificial Intelligence and Lecture Notes in Bioinformatics)*, *9564*, 68–81. https://doi.org/10.1007/978-3-319-28361-6_6

Jeon, J., Croft, W. B., Lee, J. H., & Park, S. (2006). A framework to predict the quality of answers with non-textual features. *Proceedings of the Twenty-Ninth Annual International ACM SIGIR Conference on Research and Development in Information Retrieval*, *2006*, 228–235. https://doi.org/10.1145/1148170.1148212

Julong Deynrt, D. (1989). Introduction to Grey System Theory. *The Journal of Grey System*, *1*, 1–24.

Karpinska-Krakowiak, M., & Modlinski, A. (2020). Popularity of Branded Content in Social Media. *Journal of Computer Information Systems*, *60*(4), 309–315. https://doi.org/10.1080/08874417.2018.1483212

Koufaris, M. (2002). Applying the Technology Acceptance Model and flow theory to online Consumer Behavior. *Information Systems Research*, *13*(2), 205–223. https://doi.org/10.1287/isre.13.2.205.83

Kuang, L., Huang, N., Hong, Y., & Yan, Z. (2019). Spillover Effects of Financial Incentives on Non-Incentivized User Engagement: Evidence from an Online Knowledge Exchange Platform. *Journal of Management Information Systems*, *36*(1), 289–320. https://doi.org/10.1080/07421222.2018.1550564

Kung, C. Y., & Wen, K. L. (2007). Applying Grey Relational Analysis and Grey Decision-Making to evaluate the relationship between company attributes and its financial performance-A case study of venture capital enterprises in Taiwan. *Decision Support Systems*, *43*(3), 842–852. https://doi.org/10.1016/j.dss.2006.12.012

Li, B., Jin, T., Lyu, M. R., King, I., & Mak, B. (2012). Analyzing and predicting question quality in community question answering services. *WWW'12 - Proceedings of the 21st Annual Conference on World Wide Web Companion*, 775–782. https://doi.org/10.1145/2187980.2188200

Liu, Q., Agichtein, E., Dror, G., Gabrilovich, E., Maarek, Y., Pelleg, D., & Szpektor, I. (2011). Predicting web searcher satisfaction with existing community-based answers. *SIGIR'11 - Proceedings of the*

*34th International ACM SIGIR Conference on Research and Development in Information Retrieval*, 415–424. https://doi.org/10.1145/2009916.2009974

Liu, S., Yang, Y., Cao, Y., & Xie, N. (2013). A summary on the research of GRA models. *Grey Systems: Theory and Application*, *3*(1), 7–15. https://doi.org/10.1108/20439371311293651

Liu, Y., Bian, J., & Agichtein, E. (2008). Predicting information seeker satisfaction in community question answering. *ACM SIGIR 2008 - 31st Annual International ACM SIGIR Conference on Research and Development in Information Retrieval, Proceedings*, Section 2, 483–490. https://doi.org/10.1145/1390334.1390417

Liu, Z., Niclausse, N., & Jalpa-Villanueva, C. (2001). Traffic model and performance evaluation of Web servers. *Performance Evaluation*, *46*(2–3), 77–100. https://doi.org/10.1016/S0166-5316(01)00046-3

Morán, J., Granada, E., Míguez, J. L., & Porteiro, J. (2006). Use of grey relational analysis to assess and optimize small biomass boilers. *Fuel Processing Technology*, *87*(2), 123–127. https://doi.org/10.1016/j.fuproc.2005.08.008

Ponzanelli, L., Mocci, A., Bacchelli, A., Lanza, M., & Fullerton, D. (2014). Improving low quality stack overflow post detection. *Proceedings - 30th International Conference on Software Maintenance and Evolution, ICSME 2014*, 541–544. https://doi.org/10.1109/ICSME.2014.90

Prantl, D., & Prantl, M. (2018). Website traffic measurement and rankings: competitive intelligence tools examination. *International Journal of Web Information Systems*, *14*(4), 423–437. https://doi.org/10.1108/IJWIS-01-2018-0001

Pypi. (2018). *selenium 3.141.0*. https://pypi.org/project/selenium/

Pypi. (2019). *zhihu-oauth 0.0.42*. https://pypi.org/project/zhihu-oauth/

Quoniam, L. (2011). *Competitive Inteligence 2. 0 : Organization, Innovation and Territory*. John Wiley & Sons, Incorporated. http://ebookcentral.proquest.com/lib/anu/detail.action?docID=1124683

Roy, Pradeep K., Singh, J. P., Baabdullah, A. M., Kizgin, H., & Rana, N. P. (2018). Identifying reputation collectors in community question answering (CQA) sites: Exploring the dark side of social media. *International Journal of Information Management*, *42*, 25–35. https://doi.org/10.1016/j.ijinfomgt.2018.05.003

Roy, Pradeep Kumar, Ahmad, Z., Singh, J. P., Alryalat, M. A. A., Rana, N. P., & Dwivedi, Y. K. (2018). Finding and Ranking High-Quality Answers in Community Question Answering Sites. *Global Journal of Flexible Systems Management*, *19*(1), 53–68. https://doi.org/10.1007/s40171-017-0172-6

Srba, I., & Bielikova, M. (2016). A Comprehensive Survey and Classification of Approaches. *ACM Transactions on the Web*, *10*(3), 1–63.

Stephen, G. (2019). Webometric analysis of central universities in north Eastern Region, India. A study of using Alexa internet. *Library Philosophy and Practice*, *2019*.

Strong, D. M., Lee, Y. W., & Wang, R. Y. (1997). Data quality in context DQ problems. *Communications of the ACM*, *40*(5), 103–110. http://delivery.acm.org/10.1145/260000/253804/p103-strong.pdf?ip=141.59.82.181&id=253804&acc=ACTIVE SERVICE&key=2BA2C432AB83DA15.6118EF98202DC727.4D4702B0C3E38B35.4D4702B0C3E38B35&__acm__=1519660767_f4fa7913c4dfb8d4987d460a4b761e1b%0Ahttp://portal.acm.org

Suzuki, J., Taira, H., Sasaki, Y., & Maeda, E. (2003). *Question classification using HDAG kernel*. 61–68. https://doi.org/10.3115/1119312.1119320

Taylor, Z. W., & Bicak, I. (2020). Buying search, buying students: how elite U.S. institutions employ paid search to practice academic capitalism online. *Journal of Marketing for Higher Education*, 1–26. https://doi.org/10.1080/08841241.2020.1731910

Vyas, C. (2019). Evaluating state tourism websites using Search Engine Optimization tools. *Tourism Management*, *73*, 64–70. https://doi.org/10.1016/j.tourman.2019.01.019

Wang, G., Gill, K., Mohanlal, M., Zheng, H., & Zhao, B. Y. (2013). Wisdom in the social crowd: An analysis of Quora. *WWW 2013 - Proceedings of the 22nd International Conference on World Wide Web*, 1341–1351.

Wen, K. (2004). the Grey System Analysis and Its Application. *International Journal of Computational Cognition*, *2*(1), 21–44.

Yao, Y., Tong, H., Xie, T., Akoglu, L., Xu, F., & Lu, J. (2013). *Want a Good Answer? Ask a Good Question First!* http://arxiv.org/abs/1311.6876

YueLiu, Tang, A., FeiCai, Ren, P., & Sun, Z. (2019). Multi-feature based Question–Answerer Model Matching for predicting response time in CQA. *Knowledge-Based Systems*, *182*, 104794. https://doi.org/10.1016/j.knosys.2019.06.002

Yeniterzi, R., & Callan, J. (2014). Constructing effective and efficient topic-specific authority networks for expert finding in social media. *SoMeRA 2014 - Proceedings of the 1st ACM International Workshop on Social Media Retrieval and Analysis, Co-Located with SIGIR 2014*, 45–50. https://doi.org/10.1145/2632188.2632208

Zhu, X., & Gauch, S. (2000). Incorporating quality metrics in centralized/distributed information retrieval on the World Wide Web. *SIGIR Forum (ACM Special Interest Group on Information Retrieval)*, 288–295. https://doi.org/10.1145/345508.345602